# PROGRESS ON THE DESIGN OF A PERPENDICULARLY BIASED 2ND HARMONIC CAVITY FOR THE FERMILAB BOOSTER*

R. L. Madrak[†], J. E. Dey, K. L. Duel, J. C. Kuharik, W. A. Pellico, J. S. Reid, G. Romanov, M. Slabaugh, D. Sun, C. Y. Tan, I. Terechkine,
Fermilab, Batavia, IL 60510

*Abstract*

A perpendicularly biased 2nd harmonic cavity is being designed and built for the Fermilab Booster. Its purpose is to flatten the bucket at injection and thus change the longitudinal beam distribution to decrease space charge effects. It can also help at extraction. The cavity frequency range is 76 – 106 MHz.

The power amplifier will be built using the Y567B tetrode, which is also used for the fundamental mode cavities in the Fermilab Booster. We discuss recent progress on the cavity, the biasing solenoid design and plans for testing the tuner's garnet material.

## HISTORY

Perpendicularly biased prototype cavities have been constructed in the past: at LANL, where the same cavity was later shipped to TRIUMF and tested with some modifications. SSCL was planning to use a perpendicularly biased cavity for the Low Energy Booster (LEB). Both the LANL/TRIUMF and SSCL cavities were tested at high power but neither have ever operated with beam.

The required tuning range of the FNAL 2nd Harmonic cavity (76 – 106 MHz) is even larger than that of its predecessors, (46.1 – 60.8 MHz for LANL and 47.5 – 60 MHz for SSCL [1]) so the design is challenging. Among the concerns are: 1) Achieving the required tuning range using a realistic bias magnetic field, 2) Taking into account higher local permeability and heating of the garnet in areas with low magnetic field, 3) Keeping the magnetic field in the tuner as uniform as possible, 4) Transfer and removal of heat without the use of toxic materials (i.e. BeO), or those that would generate mixed waste (oil), 5) Including the power tetrode in the RF model to take into account the impact of its output capacitance and strong coupling to the cavity on the tuning range, 6) Avoiding air bubbles (which could cause sparking) in the adhesive used to fill any air gaps in the tuner, and 7) Choosing the adhesives and assembly techniques to minimize RF losses in this adhesive without compromising heat transfer.

## CAVITY DESIGN

The tuner is constructed using a stack of rings of garnet (National Magnetics AL-800), for tuning, and alumina, to transfer the heat to the outer and inner surfaces, where it will be removed by water cooling. The required gap voltage in the cavity to improve capture during injection (75 – 82 MHz) is $V_{\text{peak}} = 100$ kV. The cavity can only be useful at transition if two are available because the required voltage exceeds 100 kV. For extraction, the required voltage is smaller. In the present CST model of the cavity, the shunt impedances ($R_{sh} = V^2/P$) are 192 kΩ at 76 MHz and 361 kΩ at 106 MHz. The design has converged regarding issues 1 - 6 enumerated above. Solutions to such problems, such as the use of magnetic shimming to make the magnetic field in the tuner more uniform, are discussed in previous proceedings including [2].

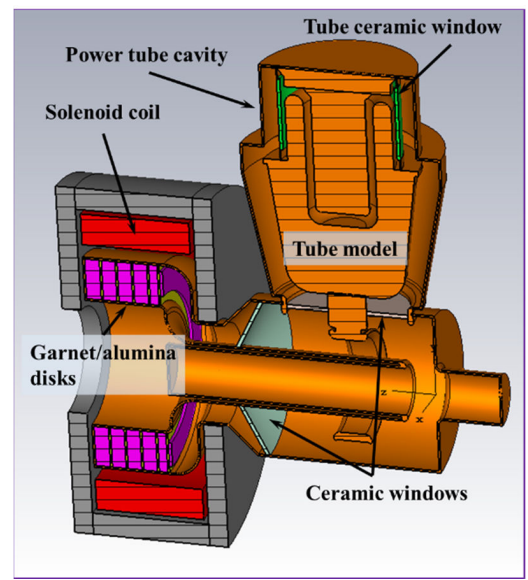

Figure 1: The present cavity design

Since the magnetic field in the garnet rings of the tuner is not uniform, it is critical to know the static permeability $\mu(B)$ and magnetic loss tangent for any bias magnetic field level within the tuner. As the vendor does not provide such data, it was necessary to measure these quantities in material samples. These first measurements are discussed in [3].

The size of the power tube and various mechanical requirements made the power tube cavity comparable in size and stored energy with the accelerating cavity itself. Implementation of the initial design of the power tube cavity in the cavity design dramatically lowered the operating frequency, and reduced the tuning range and overall effectiveness. Our attempts to take the power tube into account using simplified models with lumped elements to find a solution were not entirely successful. Progress was made after we developed a power tube model as close as possible

___

to the real one, though it pushed us to the limits of our computing capabilities. The "realistic" tube model was verified during the power amplifier tests [2] and then applied to the cavity design. With the use of this "brute force" approach we developed a special socket for the power tube to reduce the electrical length of the tube cavity, optimized its shape and the shape of the internal "shroud" to reduce stored energy and restore tuning range (see Fig. 2). Increasing the frequency at the high end of the tuning curve is especially important, because it significantly reduces the required maximum bias current. Also, the simulations assured us that the accelerating voltage of 100 kV now is real over the entire frequency range. Finally, incorporating a realistic model of the tube also allowed us to verify that the impedance seen by the tube is one that it can drive.

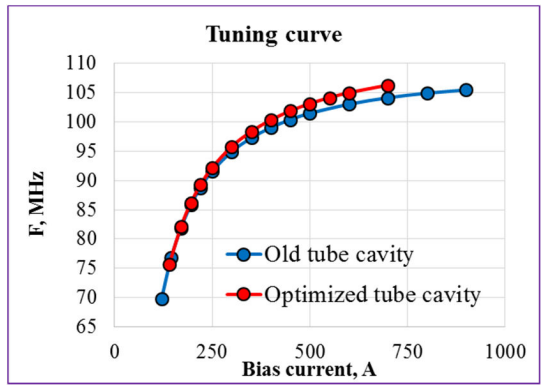

Figure 2: Tuning curves before and after optimization of the power tube cavity.

## BIAS SYSTEM DESIGN

The cavity frequency sweep is achieved by changing the bias magnetic field in the tuner. The repetition rate of the booster is 15 Hz; within each 33 ms acceleration cycle the second harmonic cavity will be used twice for 3 ms intervals, at injection and transition (or for 1 ms at extraction).

The relatively fast frequency change at injection requires a corresponding change in the bias field, so the flux return of the magnetic system must be designed taking into account the magnetic saturation and the impact of the eddy currents. Lack of longitudinal space due to the presence of a bulky tetrode (Fig. 1) forces us to use a flux return assembled using 0.025" thick M15 silicon steel laminations.

A cross-section of the magnetic bias system with the garnet rings of the RF tuner, which heavily impact the uniformity of the bias magnetic field, is shown in Fig. 3.

The coil is wound using standard 0.41" square copper wire with a Ø0.229" hole for cooling water. Geometrically, 60 turns of the wire can fit in the window of the flux return, but a portion of the longitudinal space in the winding will be occupied by current leads, which come out radially through the gaps between the blocks of the flux return.

Cooling piping of the RF part of the tuner, which removes ~2 kW of average power deposited in the garnet material (see [4] for details), will also be taken out radially through the space occupied by the coil and openings in the flux return. Taking into account these restrictions, it is assumed that the bias coil will have ~50 turns.

The power loss in the winding was evaluated given the current rise required for the desired frequency ramp (Fig. 4). The average power dissipated as heat can be as high as 6.5 kW; the corresponding cooling provision requires four parallel cooling circuits – one per layer of the coil.

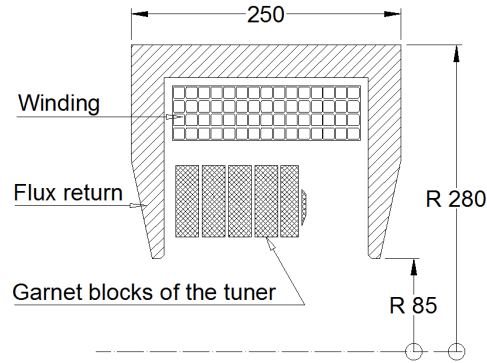

Figure 3: Schematic geometry of the bias system. Dimensions are in mm.

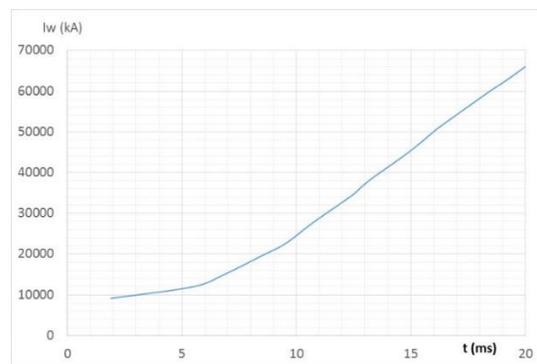

Figure 4: Required bias current ramp during the active part of the accelerating cycle.

To avoid saturation in the pole part of the flux return, it must be assembled with a sufficiently high surface density. This also helps in avoiding undesirable variation of the magnetic field on the surface of the garnet material due to the gaps between the laminated pole blocks.

With a total current $I \cdot w$ = 62.5 kA-turns, which corresponds to the end of the transition cycle, the current in the coil is 1250 A. At this current the maximum flux density in the steel reaches 1.8 T; although this is acceptable for the chosen material, making the pole parts thicker may reduce power loss due to hysteresis and eddy currents.

At the frequency of 75.73 MHz, $I \cdot w$ = 7.4 kA-turns, which corresponds to an excitation current of 148 A. At this current, the maximum value of the permeability in the garnet is ~12.5 – much larger than the average value of 3.5, which determines the frequency. Large values of the maximum permeability are associated with a corresponding low level of local magnetic field and high density of RF loss. A specially shaped garnet ring shown in Fig. 5 helps keep the maximum value of the permeability under control;

without it, the field can be too close to the gyromagnetic resonance value of ~32 Oe at 75 MHz.

The concept design of the system will be iteratively finalized following the development in the cavity design.

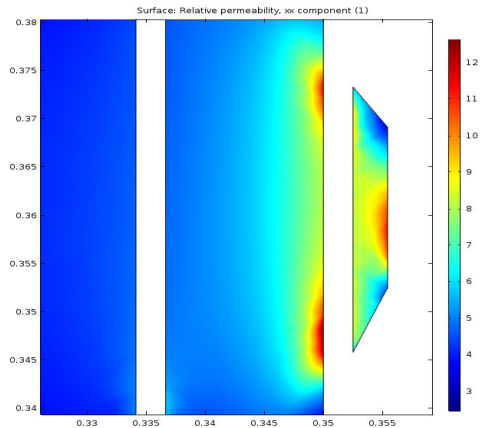

Figure 5: Permeability in the garnet material at 148 A.

## GARNET MATERIAL PROPERTIES

Each garnet ring in the tuner will be assembled using eight 45° sectors. To limit RF loss in the garnet, it is essential to have similar RF properties (resonance linewidth and saturation magnetization) in all sectors of each ring. The resonance linewidth must be kept as narrow as possible. Nevertheless, as the RF properties of the material depend on many factors, such as the purity and the proportions of the components in the mix, size of the grains, uniformity of the mix, temperature and its gradients during firing, etc., variations in the linewidth parameter are inevitable. To have some control over the local properties of the garnet rings, parameters of each sector will be measured using "witness" samples cut from the garnet material bricks after firing. Based on the measurement results, a choice will be made on how to combine the sectors in each ring. After the rings are assembled by epoxying the sectors together, the final machining will be done and RF properties of each assembly measured in a specially designed test setup. This setup consists of an RF test cavity loaded with one assembled garnet ring and installed in a gap of a special magnet that generates a sufficiently uniform magnetic field in the ring. Fig. 6 shows schematically the design of the setup.

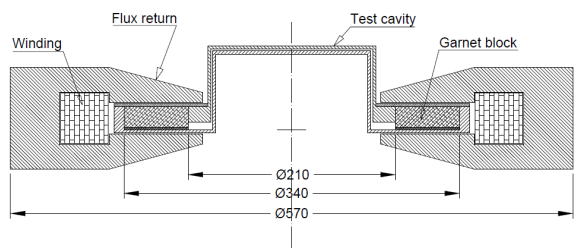

Figure 6: Setup for the measurement of the RF properties of assembled garnet rings. Dimensions are in mm.

The range of the magnetic field in the magnet and the design of the cavity are adjusted so that the properties can be measured from ~75 MHz to ~110 MHz. One of major requirements for the design is reproducibility of the measurement results after the magnet and the cavity are disassembled and reassembled to install a new garnet ring.

## EPOXY LOSS TANGENT

Stycast 2850FT with Catalyst 9 was chosen to assemble the garnet rings and to join them to the alumina because of its relatively high thermal conductivity (1.25 W/m-K). As the manufacturer specifies the dielectric constant and dissipation factor only at 1 MHz ($\varepsilon$ = 5.01 and $\tan \delta_\varepsilon$ = 0.028), it was necessary to measure the loss factor in the frequency range of the harmonic cavity. A 76 MHz quarter wave resonator was constructed from three pieces of standard 3-1/8" transmission line. A 2" thick ring of epoxy with OD and ID of ~3" and 1.375" respectively, was made to fit into the high electric field end of the resonator. The quality factor Q was measured with and without the epoxy sample. Without epoxy, it was 1628, though the simulation predicted 2342. The difference is presumably due to imperfect contact at joints, but may be approximated by changing the conductivity of the whole device. The conductivity of copper in the simulation was scaled down until the measured and simulated values of Q with no epoxy agreed. An analytical approximation of Q with epoxy resulted in $\tan \delta_\varepsilon$ = 0.017. This value was then used in the simulation with epoxy, in which case the measured and simulated values were 235 and 230. Thus the conclusion from this measurement is $\tan \delta_\varepsilon$ = 0.017, substantially smaller than the vendor's value of 0.028 at 1 MHz. As a check of the sensitivity, the value of $\tan \delta_\varepsilon$ in the simulation was increased to 0.03, in which case the simulation predicted Q = 138.

## CONCLUSION

Progress on a perpendicularly biased 2nd harmonic cavity was presented. In particular, cavity design issues were summarized and the present design was discussed. The design for the bias system was outlined, and measurements of the loss factor for the epoxy to be used were presented.

The authors are satisfied with the cavity design. Mechanical design and manufacturing has begun.